\documentclass{sig-alternate-05-2015}

\usepackage{verbatim}
\usepackage{amsmath}
\usepackage{tensor}
\usepackage{ctable}
\usepackage{multicol}
\usepackage{lipsum}
\usepackage{gensymb}
\usepackage{float}
\usepackage{caption}
\usepackage{subcaption}
\usepackage{hyperref}
\usepackage{url}
\usepackage{authblk}

\begin{document}

\CopyrightYear{2017}
\setcopyright{acmcopyright}
%\conferenceinfo{ACII '17,}{ 23-26 October, 2017, San Antonio TX,
%USA}
%\isbn{978-1-4503-4337-4/16/06}\acmPrice{\$15.00}
%\doi{http://dx.doi.org/10.1145/2910674.2910698}

% TITLE
\title{Medical Image Watermarking using 2D-DWT with Enhanced security and capacity}

\author{
\textbf{Ali Sharifara, and Amir Ghaderi} \\
Department of Computer Science and Engineering\\
University of Texas at Arlington\\
Arlington, TX 76010\\
a.sharifara@uta.edu, and amir.ghaderi@mavs.uta.edu \\ 
}

\maketitle
%------------------------------------------------------------
% ABSTRACT
%------------------------------------------------------------
\begin{abstract}
Teleradiology enables medical images to be transferred over the computer networks for many purposes including clinical interpretation, diagnosis, archive, etc. In telemedicine, medical images can be manipulated while transferring. In addition, medical information security requirements are specified by the legislative rules, and concerned entities must adhere to them. In this research, we propose a new scheme based on 2-dimensional Discrete Wavelet Transform (2D DWT) to improve the robustness and authentication of medical images. In addition, the current research improves security and capacity of watermarking using encryption and compression in medical images. The evaluation is performed on the personal dataset, which contains 194 CTI and 68 MRI cases.

\end{abstract}

% Don't do any change here
%\ccsdesc[300]{Computing methodologies~Vision for robotics}
%\ccsdesc[300]{Computing methodologies~Robotic planning}
%\ccsdesc[300]{Computing methodologies~Biometrics}
%\printccsdesc
%------------------------------------------------------------
% PAPER STARTS HERE
%------------------------------------------------------------

\section{Introduction}

Over the last decades, the development of technology has been increased especially in the field of computer networks. Digital multimedia can be stored efficiently with a high quality, and they can be manipulated easily by using computers. Moreover, digital data can be transferred in a rapid and low-cost way through data communication networks without losing their qualities.
Digital image watermarking is a special class of steganography, which is the ability of hiding a secret message in a carrier message \cite{1430770}. This is a method of embedding a digital code into a cover image without changing the image format or size and keeping the visible quality and information about the biomedical image intact \cite{4353631} . Furthermore, messages can be in the format of images, or ASCII code such as text files, or numbers \cite{5607625}.
Digital medical images as one type of multimedia can take advantage of using computer networks to send and share throughout the world. In other words, the development of the medical information scheme has changed in terms of the way of recording, method of access and distribution of patient information. Medical images are taken and stored as the EPR (Electronic Patient Record) for future access and this is quite conceivable that the digital format is simple to edit and manipulate without any degradation in image. 

Generally, there are huge archive of medical images in hospitals for clinical and diagnostic purposes and the images must be protected against malicious users \cite{Dragan:2012:RRP:2263221.2263537}. Hence, authentication of the medical images such as X-ray, MRI, Ultrasound, etc. must be secured by watermarking, in accordance with an invisible watermark related to the host image is embedded in the host image. Hence, the physicians would thus be able to proceed for the authentication phase as soon as the image is retrieved from the database. This will include the extraction of the secret message embedded in the medical image. The physician would be confident enough which some manipulations have been done when the extraction procedure of the watermark has been failed.  

\section{Related Work} 
Over the last years, exchanging the medical images for patients between hospitals has become an ordinary practice, due to the development of technologies in the areas of communications and computer networks \cite{5383112}. These medical images are exchanged for a number of reasons, which are:

\begin{itemize}
  \item Teleconferences between clinicians
  \item  for distant learning of medical personnel
  \item  Interdisciplinary exchange between physicians and radiologists for consultative purpose or to discuss diagnostic and therapeutic measures.
\end{itemize}

Medical Image Watermarking (MW) is a particular subcategory of image watermarking in the sense that the images have special requirements. One of the most significant requirements in medical image watermarking is that the image must not suffer any degradation that will modify the content of images. In general, images are required to remain intact to reach this with no visible alteration to their original formats \cite{Das:2013:EMM:2506569.2506772}. Digital image watermarking is a new approach, which is greatly suitable for archiving based applications such as medical, military, etc. The secret embedding of the watermark signal, no matter how much invisible is, it may be able to cause degradation of the image quality. Therefore, reversible watermarking is applied, which is able to conquer this drawback by applying a mechanism, which can retrieve the exact original image after the watermark has been successfully extracted.

The traditional approaches such as cryptography also can achieve this reversibility operation, but the basic defect is the loss of semantic information of the host image, for example, after encryption the medical image may not be visible or understandable, which is not the case in watermarking.
The medical watermarked images may be altered either intentionally or accidentally. The watermarking system should be robust enough to detect and extract the watermark similar to the original one \cite{1545959}. Different types of distortions, which are known as attacks, can be performed to degrade the image quality. The distortions are limited to those factors which do not produce excessive degradations in the image otherwise the transformed object would be unusable. The distortions also introduce degradation in the performance of the watermark extraction algorithm. To verify the robustness of the methods or a combination of the methods, an attack is performed intentionally on a watermarked document in order to destroy or degrade the quality of the hidden watermark.
The compression is a common attack which usually occurred on  data that's transferred via a network (either secure network or not) and the data is often compressed by using the JPEG, as high quality images like BMP images are often converted to JPEG images to reduce their size \cite{Ji:2013:DAR:2450559.2450635}. Another type of attack is removed or shuffling of blocks. In images, rows or columns of pixels may be removed or shuffled without a clear degradation in the quality of the image. All of the statements, which are mentioned above, may render an existing watermark undetectable. Salt and pepper noise is another type of attack that replaces the intensity levels of some of the pixels of an image resulting in loss of information from those pixels. Some of the popular attacks mentioned here might be intentional or unintentional, depending on the application.

\section{Proposed Method}
Image processing techniques have played important role in the past decades in the field of medical sciences for diagnosis and treatment purposes. Generally, a medical image can be divided into region of interest (ROI) and region of non-interest (RONI) \cite{1554112}. Important information regarding to diagnosis is contained in the ROI, and the rest called RONI, so its integrity must be assured. The division of ROI and RONI is a different case by case, it can be varied for different medical images, and physicians can distinguish between these two parts. We have proposed a watermarking technique to ensure the integrity of medical image, which avoids the distortion of an image in ROI by embedding the watermark information in RONI. Hence, some slight changes in the image cannot have effect in the diagnosis.

\begin{figure}[ht]
\centering
\includegraphics[width=\linewidth]{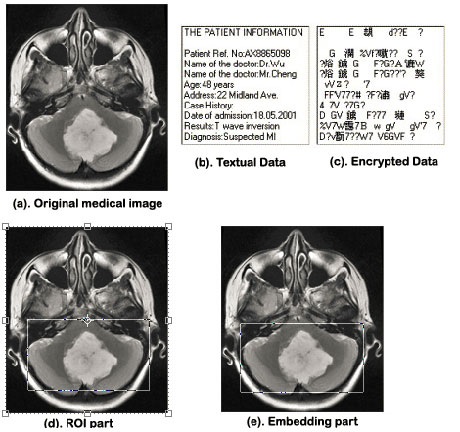}
\caption{CT scan Medical Image (a. Original medical image b. text watermark c. encrypted text watermark d. Region of Interest e. Embedding part (RONI))}
\label{fig:CTScan}
\end{figure}

The watermark contains of patient information, physician info and diagnosis result and authentication code computed using a hash function. Moreover, earlier encryption of watermark is performed to ensure inaccessibility of embedded data to the adversaries.
The algorithms and implementation of existing research have been simulated by MATLAB R2010b and C++. Figure \ref{fig:CTScan} depicts medical image (host image) in size of $ 256 \times 256 $ along with text watermarking which contains of patient information such as: patient name, name of the physician, age of patient, address of the patient, date of submission and diagnosis result.

\subsection{Process of Generating Watermark }

\textbf{Step 1}: Read watermark image, which is an 8-bits gray-scale image and can contain of some information such as logo of hospital, name of patient, name of physician etc. If the input is in different format (non- gray scale) one pre-processing step has been used to convert it into gray-scale format.

This step contains of some processing which are: 
\begin{itemize}
  \item Convert input image to gray scale.
  \item Resize watermark logo to 32x32 pixels.
\end{itemize}

Converting the watermark into binary by using a threshold, which we call it, T (usually 127) and the algorithms, converts the image by following formula:  

\begin{equation}
  T=
  \begin{cases}
    make the pixel white (1), & \text{if}\ image(x, y) > T \\
    make the pixel black (0), & \text{otherwise}
  \end{cases}
\end{equation}

Where, x and y are the row and column indicates of logo image, which in our work for the watermark is $ 1 \leq x \leq 32  and 1 \leq y \leq 32 $. Then we convert it into the vector and we call it $W_1$  which is:

\begin{equation}
  W_1 = {v1(i)|v1(i) \in [0,1], 1 \leq 1024}.
\end{equation}

\textbf {Step 2}: Read a text file which can contain of patient information, hospital info and etc and convert it into ASCII code and then convert ASCII code into a binary code by using following formula:

\begin{equation}
  W_2 = {v2(i)|v2(i) \in [0,1], 1 \leq i \leq L}.
\end{equation} 
\\
Where, L is the length of bits and in our model, we have used 512 character of patient, which is become: $512\times8=4096$ bits.

\textbf{Step 3}: Concatenate of two watermarks $W_1, W_2$ to make W with the length of M. 

\begin{equation}
  W = {W(i)|W(i) \in [0,1], 1 \leq i \leq M}.
\end{equation}

\subsection{Process of Embedding the Watermark}
\textbf{Step 1}: Read two input images, which are medical image (host image) and the watermark. Medical image is an 8-bits gray-scale image and watermark is a grayscale image, which can contain of some information such as logo of hospital, name of patient, name of physician etc. If the inputs are in different format (non- gray scale) one pre-processing step has been used to convert them into gray-scale format. This step contains of some processing which are: \\
\begin{itemize}
  \item  Convert input image to gray scale.
  \item Resize cover image to $ 256 \times 256 $ pixels. 
\end{itemize} 

\textbf{Step 2}: Dividing the received medical image into ROI and RONI. (Physicians can decide which part is ROI and which part is RONI). \\
\textbf{Step 3}: Extracting and saving dimensions of ROI into two variables. This makes our method be flexible and compatible with any size of medical images. \\
\textbf{Step 4} The RONI is divided into 6x1 pixels blocks. After the original LSBs is compressed using RLE, each resulting RLE package will be embedded in a block in RONI. \\
\textbf{Step 5}: The ROI in medical image is transferred to wavelet domain by using Discrete Wavelet Transform (DWT). For this purpose, we have used "Haar" transform application. The ROI decomposed into 4 sub-domains which are HH, HL, LH and LL.  This decomposition happens according to the different frequencies of the cover image. \\
\textbf{Step 6}: Transformed medical image (ROI) into one dimension and the length of dimension will be calculated as well as size of each sub-domain.\\
\textbf{Step 7}:Determine the maximum coefficient value in each 4 sub-domains. \\
\textbf{Step 8}: The positions which binary logo can be embedded into sub-domains will be recognized and each sub domain must be between 0 and maximum value of selected domain. \\
\textbf{Step 9}: Next, for improving the robustness of our proposed method we hide more than one set of watermark into the medical image (RONI area). It can make the watermark recoverable with good quality. The hiding process in each of these domains follows a specified formula. \\
\textbf{Step 10}: Finally, the decomposed image is converted to 2D dimension and the watermark logo is embedded inside the medical image to make watermarked image. \\ 

\subsection{Increasing the embedding capacity}

\begin{figure}[ht]
\centering
\includegraphics[width=\linewidth]{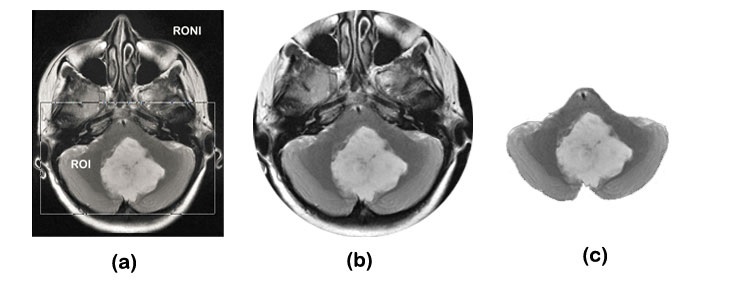}
\caption{(a) ROI with using rectangle (b) ROI with using square (c) Proposed Algorithm}
\label{fig:ROI}
\end{figure}

The capacity of watermarked image depends on the medical image as well as method of selecting RONI. By isolating the brain CT scan with a technique, one may increase embedding capacity for the watermark insertion. By taking the square for isolating the medical image from the input CT scan image with size $256\times256$ pixels, it was necessary to take at least square of size 192x192 pixels in order to cover the whole brain as shown in Figure \ref{fig:ROI}(a). This gives $[(256\times 256)$ $(192\times192)]=28672$ pixels for embedding the watermark information. Similarly drawing the ellipse as shown in Figure \ref{fig:ROI}(b) gives $[(256\times256) 33749] = 31787$ pixels as embedding capacity. With the proposed scheme shown in Figure \ref{fig:ROI} $(c), [(256\times256) 17114] = 48422 $ pixels are obtained for embedding watermark information.

\begin{figure*}[ht]
\centering
\includegraphics[width=\linewidth]{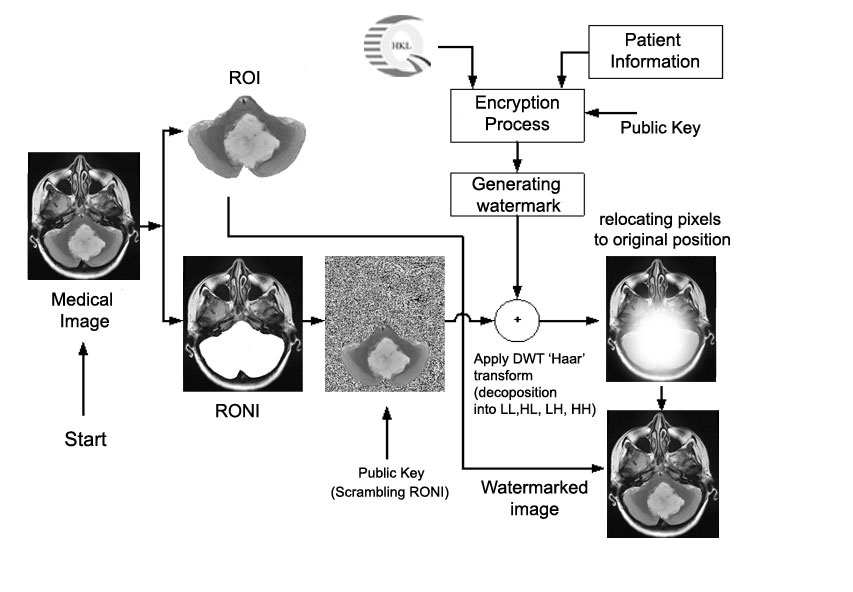}
\caption{Block diagram of Embedding process}
\label{fig:Embedding}
\end{figure*}

Figure \ref{fig:Embedding} depicts the block diagram of embedding watermark in medical image. by this assumption that there are two types of watermarks to embed which are patient information and hospital logo which uses text and image watermark.

\subsection{Extracting Process of Watermark }

As far as the proposed method is non-blind watermarking, we postulate that in the recovery time, the watermark is also available. Hence, in the process of recovery we first take the original image that has been used to hide the information. Along with that we also send the receiver of the message, 2 keys which essentially act as private keys. These keys are necessary to decrypt and to extract the encrypted embedded data. The steps in the extracting process are as follows:\\
\textbf{Step 1}: The Medical Image and Watermarked image are taken from the input. \\
\textbf{Step 2} Secondly, the first level decomposition of the watermarked image and medical image will be extracted by using DWT.\\
\textbf{Step 3}The watermarked image and medical image are reshaped into one dimension.\\
\textbf{Step 4}: The two input keys are taken from the user equal to the dimension of the logo. \\
\textbf{Step 5}: After that we have to determine the maximum coefficient values of the original cover image. \\
\textbf{Step 6}: Those positions are then found that were used to hide the logo for each of the 4 decompositions. \\
\textbf{Step 7}: Positional sets for different sets of the logo from each decomposition are extracted. \\
\textbf{Step 8}: Finally, different sets of logo are recovered from each of the sub bands using bit majority algorithm and the final logo is constructed from the different recovered sets.

\begin{figure}[ht]
\centering
\includegraphics[width=\linewidth]{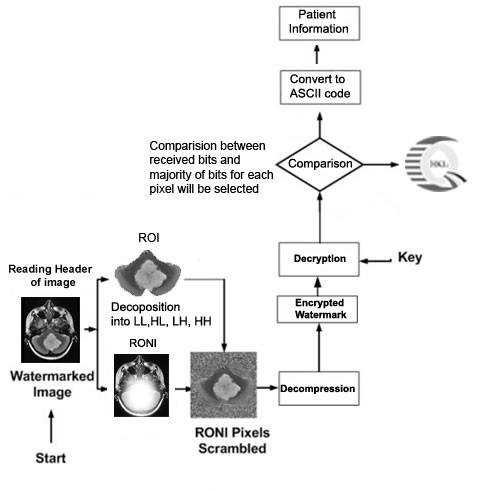}
\caption{Block diagram of extracting process.}
\label{fig:Extracting}
\end{figure}

Figure \ref{fig:Extracting} shows the block diagram of extracting process which is the reverse of embedding process.

\section{Result and Analysis} 

To evaluate the amount of quality between the medical watermarked images and the original medical image, different kinds of image quality metrics have been applied. In the current study, we have applied structural similarity index measure (SSIM) and peak signal-to-noise ratio (PSNR) in our dataset, contains 194 CTI and 68 MRI cases.

\subsection{Structural Similarity Index Measure (SSIM)}

The SSIM has been used to measure the similarity  between two images (watermarked and original image) \cite{Thung:2012:CIQ:2142120.2142201}.

\begin{equation}
SSIM(x,y)=\frac{(2\mu_x \mu_y + c_1)(2\sigma_{xy} + c_2)}{(\sigma_x^2 + \sigma_y^2 + c_1)(\sigma_x^2 + \sigma_y^2 + c_2)}
\end{equation}

Where, \\
$ \mu_x $ is the average of x \\
$ \mu_y $ is the average of y \\
$ \sigma^2_x $ is the variance of x \\
$ \sigma^2_y $ is the variance of y \\
$ \sigma_{xy^i} $ is the covariance of x and y \\
$ L $ is the dynamic range of pixel-values \\
$ C_1 = (K_1k)^2, C_2 = (K_2L)^2 $ \\
$ K_1 = 0.01 $ and $ K_2 = 0.03 $ by default \\

The result of SSIM is a decimal value between -1 and 1. The value of 1 is only achievable when two input images (in our case, watermarked and host images) are matched.

\subsection{Peak Signal-to-Noise Ratio (PSNR)}

It is the ratio, which measures the influential corrupting noise and signals effectiveness' on fidelity of its representation, which have a determined relation \cite{Hasnaoui:2014:MQV:2574577.2574697}. PSNR is usually declared in terms of dB. It is most simply defined through the mean square error (MSE) which is used for $ 2 m \times n $ images.

\begin{equation}
\begin{split}
& MSE=\frac{1}{MN}  {\sum\limits_{n=0}^M} {\sum\limits_{m=0}^N} [I(i,j)-k(i,j)]^2 \\
& PSNR=10.log_{10} \Big(\frac{MAX_1^2}{MSE}\Big) \\
&  \qquad \quad = 20.log_{10}\Big(\frac{MAX_1}{\sqrt[]{MSE}}\Big)
\end{split}
\end{equation}

Where, $MAX_1$ is the maximum possible pixel value. In this example we have used gray-scale image and the maximum for 8 bit is 255. Different types of medical images have been watermarked by using the proposed watermarking system, which were presented in Sections 3. Table 1 shows PSNR and SSIM for the proposed method and also Table 2 depicts the percentage of watermark damaged through the extraction process.

\begin{table}[h!]
\centering
\begin{tabular}{|c|c|c|} 
 \hline
\textbf{Method} & \textbf{SSIM} & \textbf{PSNR} \\ [0.5ex] 
 \hline
 Proposed Method & 0.9891 & 42.453  \\ 
  \hline
\end{tabular}
\caption{PSNR and SSIM for the proposed method}
\label{table:PSNR}
\end{table}

\begin{table}[h!]
\centering
\begin{tabular}{|c|c|c|c|c|} 
 \hline
\textbf{Image} 
&
\textbf{capacity} & \textbf{ 
\begin{tabular}{@{}c@{}}Max \\ Capacity\end{tabular}} 
& 
\textbf{\begin{tabular}{@{}c@{}}Regions \\  \end{tabular}}
& 
\textbf{\begin{tabular}{@{}c@{}} Damaged \\ \end{tabular}}
  \\
 \hline
 CTI & 3999 & 8192 & 48.77 & 0.1244 \\ 
  \hline
 MRI & 634 & 2048 & 31.78 & 0.1556 \\ 
  \hline
\end{tabular}
\caption{Percentage of watermark damaged through the extraction process}
\label{table:Damage}
\end{table}

\section{Conclusion}

The present paper proposed a novel medical image watermarking scheme for medical images, which embeds watermark and patient information in medical images. The scheme is proposed to preserve the visual integrity of medical images, which must not be compromised by watermarking. The proposed scheme has been used compression method to increase the size of watermark and also encryption to increase the security of medical images. Moreover, the current method combines image and text data as a watermark and then embeds data to RONI part of medical images. 

\bibliographystyle{abbrv}
\bibliography{references}
 % sigproc.bib is the name of the Bibliography in this case
 % NEED REFERENCES TO CODE LIBRARIES USED IN PROJECT: OpenCV, SciPy, NumPy, 
\end{document}